\documentclass[%
 reprint,
 amsmath,amssymb,
 aps,superscriptaddress
]{revtex4-2}

\usepackage{graphicx}
\usepackage{dcolumn}
\usepackage{bm}
\usepackage[normalem]{ulem}
\usepackage{xcolor}
\usepackage{mathrsfs}
\usepackage{inputenc}
\usepackage[slantedGreek]{mathpazo}
\usepackage{hyperref}
\usepackage{float}
\begin{document}
\newcommand{\rb}[1]{\textcolor{red}{\it#1}}
\newcommand{\rbout}[1]{\textcolor{red}{\sout{#1}}}

\preprint{APS/123-QED}

\title {Ab initio spectroscopic studies of AlF and AlCl molecules}
\author{R. Bala}
\email{balar180@gmail.com}
\author{V. S. Prasannaa}
\author{D. Chakravarti}
\author{D. Mukherjee}
\author{B. P. Das}
\affiliation{Centre for Quantum Engineering, Research and Education, TCG CREST, Salt Lake, Kolkata 700091, India\\}
\date{\today}

\begin{abstract}
In this work, we report results from our extensive spectroscopic study on AlF and AlCl molecules, keeping in mind potential laboratory as well as astrophysical applications. We carry out detailed electronic structure calculations in both the molecules, including obtaining the potential energy surfaces of the $X^1\Sigma$ ground electronic state and some of the relevant low-lying excited electronic states belonging to $\Sigma$ and $\Pi$ symmetries. This is followed by evaluating spectroscopic constants and  molecular properties such as electric dipole moments and electric quadrupole moments. Throughout, we employ the multi-reference configuration interaction method and work with high-quality quadruple zeta basis sets, keeping in mind the need for precise results. Further, transition dipole moments between the ground electronic state and singlet excited states are also studied. The relevant vibrational parameters are computed by solving the vibrational Schr\"odinger equation. Subsequently, the vibrational energy spacings and transition dipole moments between the vibrational levels belonging to the same electronic states are used to evaluate the spontaneous and black-body radiation induced transition rates, followed by computing lifetimes. Finally, the energy differences between rotational levels belonging to different vibrational levels and within an electronic state as well as Einstein coefficients are reported. \\
\begin{description}
\item[Keywords]
{\it ab initio} calculations, spectroscopic constants, molecular properties, potential energy curves, dipole moment, quadrupole moment, transition dipole moment, transition rate, lifetime, vibrational and rotational transitions.
\end{description}
\end{abstract}
\maketitle

\section{\label{sec:level1}INTRODUCTION\protect\\ }
\label{section-1}

The field of cold and ultracold molecules is one of the fastest growing areas of research due to its scope in a wide range of fields, including quantum chemistry\,\cite{Bell_2009}, fundamental physics\,\cite{Zelevinsky_2022}, quantum computing\,\cite{Demille_2002, Yelin_2006}, quantum simulation\,\cite{Blackmore_2018}, controlled chemical reactions~\cite{Ospelkaus_2010, Kosicki_2017}, anisotropic- and long-range dipole-dipole interactions~\cite{Menotti_2008, Menotti_2011}, etc. An important feature that makes a molecule particularly attractive for many applications is its laser coolability. Further, extensive knowledge and careful analysis of the properties and spectroscopy of such systems could open avenues for a wider range of applications in numerous interrelated areas of research.\\

AlF and AlCl molecules have recently been proposed as promising candidates for laser cooling and trapping experiments, and efforts towards laser cooling these systems are currently underway~\cite{Rosa_2004, Yang_2016, Wan_2016, Daniel_2021, Hofsass_2021}. The rate coefficients for the formation of AlF and AlCl molecules from Al, F and Cl atoms via radiative association have been predicted by Andreazza {\it et al.}~\cite{Andreazza_2013, Andreazza_2018}. These molecules can be produced in gas phase by heating Al with AlF$_3$ or AlCl$_3$ to high temperatures~\cite{Hedderich_1993}. \\ 

These molecules are also of interest to astrophysics due to their occurrence in astrophysical environments \cite{Raja_2006,Bagare_2006}. In particular, the rotational energy levels of these molecules have been identified in the circumsteller envelope of a C-type asymptotic giant branch (AGB) star, IRC +10216~\cite{Agundez_2012}. Quite recently, AlF emission lines have been detected in five M-type AGB stars~\cite{Saberi_2022}. \\ 

 \begin{table*}[t]
\caption{\label{tab:table1}
Spectroscopic constants for the ground electronic state of the AlF molecule. The results computed in our work are shown in bold fonts. The term Exp. is used to represent experimental results.}
\begin{ruledtabular}
\begin{tabular}{cccccccc}
Method &$R_e$ (\AA) & $D_e$ (eV) & $B_e$ (cm$^{-1}$) & $\alpha_e$ (cm$^{-1}$) & $\omega_e$ (cm$^{-1}$)& $\omega_e x_e$ (cm$^{-1}$) & $\omega_e y_e$ (cm$^{-1}$)\\
\hline
\textbf {SCF} & \textbf{1.644} & \textbf{9.343} & \textbf{0.5609} & \textbf{0.00509} & \textbf{845.22} & \textbf{5.51} & \textbf{0.0243} \\
\textbf{CASSCF} & \textbf{1.669} & \textbf{9.666} & \textbf{0.5447} & \textbf{0.00519} & \textbf{802.26} & \textbf{5.09} & \textbf{0.0231}\\
\textbf{MRCI} & \textbf{1.666} & \textbf{7.206} & \textbf{0.5459} & \textbf{0.00500} & \textbf{799.49} & \textbf{4.54} & \textbf{0.0161}\\
\textbf{MRCI\,+\,Q} & \textbf{1.667} & \textbf{7.022} & \textbf{0.5433} & \textbf{0.00416} & \textbf{798.02} & \textbf{5.25} & \textbf{0.0314}\\
\hline
\textbf{Published works}\\
CASSCF \cite{Langhoff_1988}& 1.681 & 5.710 &$-$& $-$ & 773.00& $-$&$-$\\
MRCI \cite{Langhoff_1988}&1.675 & 6.760 & $-$ & $-$ & 788.00 & $-$ & $-$\\
MRCI\,+\,Q \cite{Langhoff_1988}&1.679 & 6.870 &$-$ & $-$& 777.00& $-$ & $-$\\
MRCI\,+\,Q \cite{Zhang_2021}& 1.664 & 7.036 & 0.5459 & $-$ & 796.11 & 4.45 & $-$\\
CCSD(T) \cite{Woon_2009}&1.680 & $-$ & 0.5360 &$-$ & $-$ & $-$ &$-$\\
MRCI \cite{Well_2011}& 1.638 & $-$ & $-$ & $-$ & 777.00 & $-$ & $-$\\
CAM-B3LYP \cite{Walter_2022} & 1.659 & $-$ & $-$ & $-$  & 792.87&$-$ & $-$\\
ACPF \cite{Yousefi_2018} & 1.654 & $-$ & 0.5530 & 0.00494 & 806.35 & 4.72 & $-$\\
CCSD(T) \cite{Gutsev_1999}& 1.660 & 7.010 & 0.5490 & $-$ & 803.00&$-$ &$-$\\
Exp.\cite{Hedderich_1993}& 1.654 & 6.943 & $-$ & $-$ & $-$ & $-$ & $-$\\
Exp. \cite{Lide_1965}& 1.65437 $\pm$ 0.00001& $-$ & 0.55246 & 0.00495 & $-$ &$-$ & $-$\\
Exp. \cite{NIST, Herzberg_1979}& 1.654369 &$-$ & 0.5524798 & 0.0049841 & 802.26 & 4.77 & $-$ \\
\end{tabular}
$^\dag$ All the results from literature have been rounded off (Except experimental data) up to the same decimal place to that reported in the current work. 
\end{ruledtabular}
\end{table*}

Using a jet-cooled, pulsed molecular beam, spectroscopic characterization of the lowest rotational levels in the $a^3\Pi$ state of AlF for $v = 0$ to $8$ have been reported by Walter {\it et al.}~\cite{Walter_2022}. Recently, spectroscopic measurements of AlF have been performed for energy level structures in $X^1\Sigma$, $A^1\Pi$ and $a^3\Pi$ states by Truppe {\it et al.}~\cite{Truppe_2019}. For the molecule, a high-resolution emission spectrum lying in the infrared region has been observed by Bernath {\it et al.}~\cite{Bernath_1992}, using Fourier transform infrared spectrometer. The emission and absorption band spectra between singlet ($A^1\Pi\,-\,X^1\Sigma,\,C^1\Sigma\,-\,A^1\Pi,\,D^1\Delta\,-\,A^1\Pi,\,  F^1\Pi\,-\,A^1\Pi,\,G^1\Sigma\,-\,A^1\Pi,\,F^1\Pi\,-\,B^1\Sigma$), triplet ($^3\Sigma\,-\,^3\Pi$) and intercombination electronic states ($a^3\Pi\,-\,X^1\Sigma$) fall in different spectral ranges, and have been observed and reported in literature~\cite{Barrow_1953, Rochester_1939, Hugo_1953, Hugo_1954, Broida_1976}.\\

In addition to experimental data for spectroscopic parameters of AlF molecule, there are several theoretical results reported in literature. For instance, Langhoff {\it et al.}~\cite{Langhoff_1988} have reported the spectroscopic parameters (equilibrium bond length ($R_e$), harmonic frequency ($\omega_e$), vertical transition energy ($T_e$), and dissociation energy ($D_e$)) for the ground and low-lying excited states. However, they have plotted the potential energy curves (PECs) for only $X^1\Sigma$ and $A^1\Pi$ states. The feasibility to obtain a lower Doppler temperature of around $3\,\mu K$ using spin forbidden transition $a^3\Pi\,\leftarrow\,X^1\Sigma$ has been studied in Ref.~\cite{Well_2011}. Gutsev {\it et al.}~\cite{Gutsev_1999} have studied diatomic constants ($R_e$, $\omega_e$, and rotational constant ($B_e$)) and first-order molecular properties (dipole moment ($\mu$) and $z$-component of quadrupole moment ($\Theta_{zz}$)) using coupled cluster singles and doubles with partial triples (CCSD(T)) method and atomic natural orbital (ANO) basis sets. The equilibrium bond lengths, rotational constants and components of dipole polarizabilities for the ground states of both the molecules have been reported by Woon and Herbst \cite{Woon_2009} at CCSD(T)/augmented correlation-consistent polarized valence triple zeta (aug-cc-pVTZ) level. Zhang {\it et al.}~\cite{Zhang_2021} have studied the $X^1\Sigma$ and $A^1\Pi$ states at multi-reference configuration interaction (MRCI) level of theory using AWCV5Z basis set. The ro-vibrational transitions in the electronic ground states of AlF and AlCl molecular systems have been studied in Ref.~\cite{Yousefi_2018}. \\

For AlCl,  the chemical mechanisms of laser ablation sources for the production of AlCl molecules for laser cooling have been studied by Lewis {\it et al.}~\cite{Lewis_2021}. The ground and excited states of AlCl molecule have been observed by David {\it et al.}~\cite{David_1993} using resonance-enhanced multiphoton ionization spectroscopy. High resolution emission spectra of AlCl at $20\mu m$ has been analysed by Hedderich {\it et al.}~\cite{Hedderich_1993}. The emission spectrum of $A^1\Pi - X^1\Sigma$ transition has been analyzed by Mahieu {\it et al.}~\cite{Mahieu1989} to obtain the vibrational and rotational constants for the involved electronic states. Ram {\it et al.}~\cite{Ram_1982} have studied the rotational bands of $A^1\Pi - X^1\Sigma$, $a^3\Pi - X^1\Sigma$, and $b^3\Sigma - a^3\Pi$ transitions to obtain the molecular constants for the electronic states of AlCl.\\

In theory, a three-electronic level scheme for laser cooling of AlCl molecule has been discussed in Ref.~\cite{Wan_2016}. Xu {\it et al.}~\cite{Xu_2020} have studied the line intensities of $X^1\Sigma\rightarrow A^1\Pi$  transition along with the spectroscopic parameters of $X^1\Sigma$, $a^3\Pi$, and $A^1\Pi$ electronic states. AlCl and its cationic species, AlCl$^+$ and AlCl$^{2+}$, have been studied and their ground and excited state parameters obtained by Brites {\it et al.}~\cite{Brites_2008}. The work by Yang {\it et al.}~\cite{Yang_2016} has reported on the suitability of AlCl and AlBr molecules for laser cooling using {\it ab initio} methods. \\

Most of the theoretical works discussed above do not report all the spectroscopic parameters ($R_e$, $D_e$, $B_e$, $\omega_e$, rotational-vibrational coupling constant ($\alpha_e$), anharmonic constants ($\omega_e x_e$ and $\omega_e y_e$)) for both ground and excited states. For example, the dissociation energies are not available for the excited states of AlF molecule except for $A^1\Pi$ state in Ref.~\cite{Zhang_2021}. The value of $\omega_e y_e$, for the $X^1\Sigma$ state of AlF is not reported by any work, while for AlCl, $\omega_e y_e$ is reported only in  Ref.~\cite{Brites_2008} to the best of our knowledge. The $\alpha_e$ and $\omega_e x_e$ results for excited states are reported only in Ref.~\cite{NIST}. Further, the rotational constants, $B_e$, for $A^1\Pi$, $B^1\Sigma$ and $b^3\Sigma$  electronic states are communicated in Ref.~\cite{NIST}.  \\ 
\begin{figure*}[t]
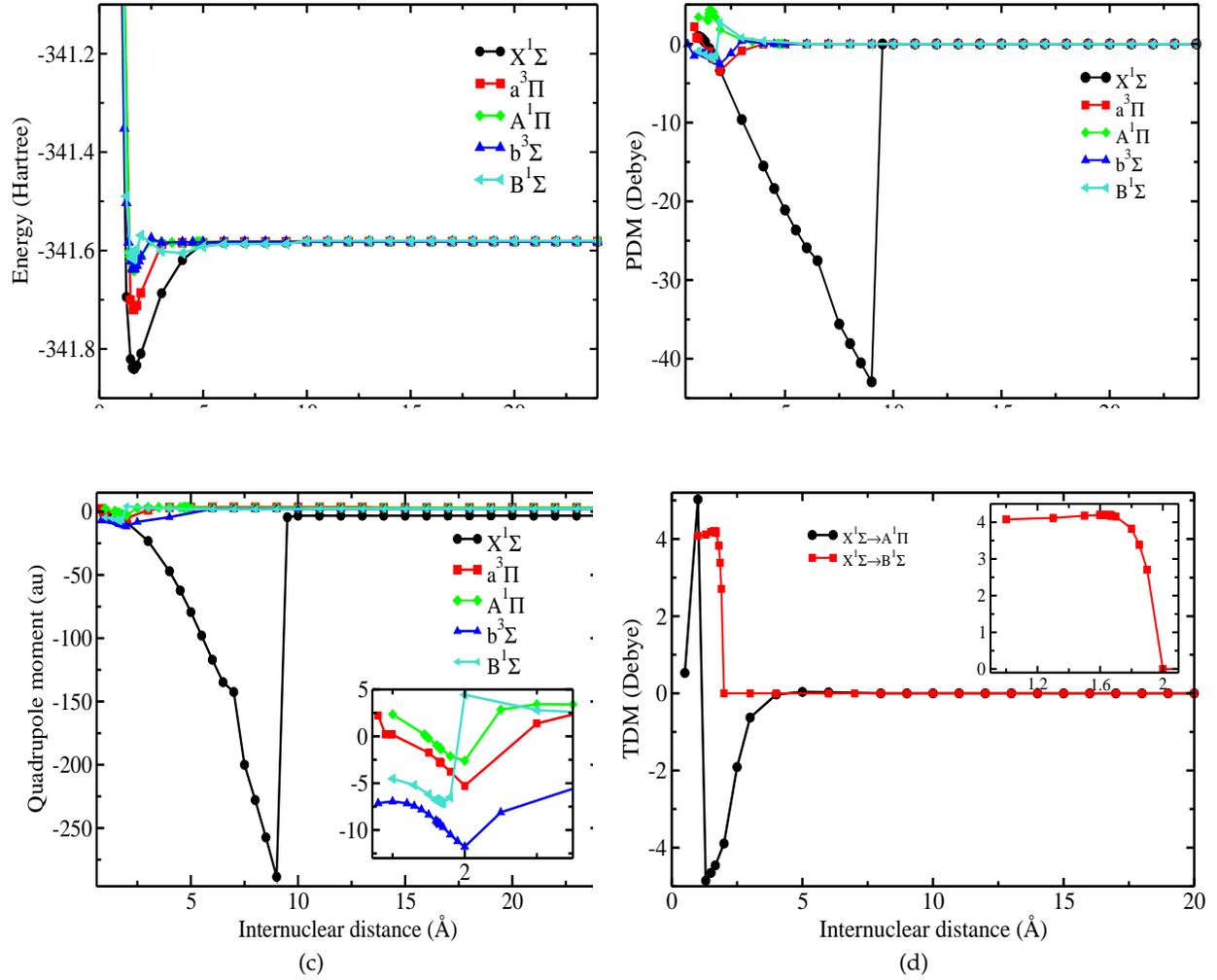

\begin{tabular}{cc}
\includegraphics[width=8.1cm, height=6.2cm]{AlF_MRCIQ_PEC_corrected.eps}\,\,&\includegraphics[width=7.9cm, height=6.2cm]{AlF_GS_ES_PDM.eps}\\
(a)&(b)\\
\,\,\includegraphics[width=7.9cm, height=6.2cm] {AlF_corrected_QM_GS_ES.eps}&  \includegraphics[width=8.0cm, height=6.2cm]{AlF_TDM.eps}\\
(c)&(d)\\
\end{tabular}
\caption{\label{fig:FIG1} (a) Potential energy curves, (b) permanent electric dipole moment curves, (c) electric quadrupole moment curves for the ground and low-lying excited electronic states, and (d) transition dipole moment curves from ground to singlet excited states (A$^1\Pi$ and B$^1\Sigma$) of AlF molecule. }
\end{figure*} 

In this work, we identify and attempt to bridge important gaps in the available data for the spectroscopic properties of these two molecular systems. We provide a comprehensive study of PECs, permanent dipole moment (PDM) curves, quadrupole moment (QM) curves, transition dipole moment (TDM) curves, and spectroscopic constants:  $R_e$, $D_e$, $B_e$, $\alpha_e$, $T_e$, $\omega_e$, $\omega_e\,x_e$ and $\omega_e\,y_e$  for not only ground state but also for low-lying excited states of AlF and AlCl molecular systems. All calculations are performed at MRCI level of theory together with the quadruple zeta (QZ) quality basis sets. It is also worth adding that no previous work has evaluated QMs for all electronic states of AlCl and excited electronic states of AlF system. Further, this study has been extended to the calculation of vibrational parameters {\it{viz.}} energy ($E_v$), vibrational-rotational constants ($B_v$), TDMs ($\mu_{v',v"}$), spontaneous and black-body radiation (BBR) induced transition rates ($\Gamma_{spon}$ and $\Gamma_{BBR}$), and lifetimes ($\tau$). We have also studied the rotational parameters in the vibrational levels of each electronic state. Furthermore, there are some discrepancies in the values of some parameters and nature of the states, which we tried to address in the current work. The aim of this work is to compute the diatomic constants, molecular properties, and spectroscopic parameters for ro-vibrational levels of ground- and low-lying excited electronic states states of AlF and AlCl molecular systems at a uniform level of theory, basis sets and configuration space.\\

The manuscript is structured as follows: Section~\ref{section-1} provides an introduction and literature review that then leads us to the motivation of the current work. The details of our calculations are described in Section~\ref{section-2}, while the obtained results are discussed in Section~\ref{section-3}. Finally, a summary of our work is given in Section~\ref{section-4}. 
 
\begin{table*}[htbp]
\begin{ruledtabular}
\begin{center}
\caption{\label{tab:table2}
Spectroscopic constants for the excited electronic states of the AlF molecule. The results computed in the present work are given in bold fonts. The abbreviation Exp. is used for the experimental results. }
\begin{tabular}{cccccccccc}
 Method & $R_e$ (\AA) &$D_e$ (eV) &$B_e$ (cm$^{-1}$) & $\alpha_e$ (cm$^{-1}$) &$T_e$ (cm$^{-1}$) &$\omega_e$ (cm$^{-1}$)&  $\omega_e x_e$ (cm$^{-1}$) &$\omega_e y_e$ (cm$^{-1}$)\\
\hline
\multicolumn{10}{c}{\textbf{$a^3\Pi$}}\\
\hline
 \textbf{MRCI} & \textbf{1.662} & \textbf{3.932} & \textbf{0.5450} & \textbf{0.0032} & \textbf{26593.46} & \textbf{826.21} & \textbf{5.31} & \textbf{0.0150}\\
\textbf{MRCI\,+\,Q} & \textbf{1.663} & \textbf{3.743} & \textbf{0.5442} &\textbf{0.0032} & \textbf{26516.33} &\textbf{825.80} & \textbf{5.43} & \textbf{0.0149} \\
 MRCI~\cite{Well_2011} & 1.631 & $-$ & $-$ & $-$& 27607.01 & 845.00 & $-$ & $-$ & \\
 CAM-B3LYP~\cite{Walter_2022} & 1.655 & $-$ &$-$ & $-$ & 25587.86 & 818.72 &$-$ & $-$ &\\
 MRCI~\cite{Langhoff_1988} & 1.659 & $-$ &$-$  & $-$ &  26035.00 & 855.00 & $-$ &$-$ &\\
 MRCI\,+\,Q~\cite{Langhoff_1988} & 1.664 &$-$  & $-$ & $- $ & 26480.00 & 845.00 & $-$ & $-$ &\\
 CCSD(T)~\cite{Gutsev_1999}& 1.653 & $-$ & 0.5530 & $-$ & $-$ & 830.00 & $-$ &$-$ & \\
 Fitting$^*$~\cite{Walter_2022} & $-$ & $-$ & 0.5574  & 0.0048 & 27239.45 & 830.28 & 4.64 & 0.0123 &\\
 Exp.NIST \cite{NIST, Herzberg_1979}& 1.6476& $-$ & 0.5570 & 0.00453&27241 & 827.8& 3.9 & $-$ &\\
\hline
\multicolumn{10}{c}{\textbf{$A^1\Pi$}}\\
\hline
\textbf{MRCI} & \textbf{1.661} & \textbf{1.715} & \textbf{0.5480} & \textbf{0.0051} & \textbf{44455.44} & \textbf{809.03} & \textbf{6.89} & \textbf{0.0089} \\
\textbf{MRCI\,+\,Q} & \textbf{1.663} & \textbf{1.591} & \textbf{0.5474} & \textbf{0.0049} & \textbf{43847.36} & \textbf{804.32} & \textbf{5.70} & \textbf{0.1946} \\
MRCI~\cite{Zhang_2021} & 1.658 & 1.591 & 0.5501 & $-$ & 43915.18 & 804.25 & 5.59 & $-$ & \\
 MRCI~\cite{Well_2011} & 1.638 & $-$ & $-$& $-$ & 44924.57 & 828.00 & $-$ & $-$&\\
MRCI~\cite{Langhoff_1988} & 1.655 & $-$ &$-$ & $-$& 44873.00& 845.00 &$-$  & $-$ &\\
MRCI\,+\,Q~\cite{Langhoff_1988} &1.664 & $-$& $-$ & $-$& 44211.00 & 828.00 & $-$ &$-$ & \\
Exp.~NIST \cite{NIST, Herzberg_1979}& 1.6485 & $-$ & 0.55640 & 0.00534 & 43949.2 & 803.90 & 5.99 & 0.050 & \\
\hline
\multicolumn{10}{c}{$b^3\Sigma$}\\
\hline
\textbf{MRCI} & \textbf{1.656}& \textbf{1.654} & \textbf{0.5104} & \textbf{0.0061} & \textbf{44764.65} & \textbf{787.20} & \textbf{6.42} & \textbf{0.2110} \\
\textbf{MRCI\,+\,Q} & \textbf{1.660} & \textbf{1.514} & \textbf{0.5501} & \textbf{0.0053} & \textbf{44438.34} & \textbf{791.26} & \textbf{8.81} & \textbf{0.0516} \\
MRCI~\cite{Well_2011} & 1.615 &$-$ & $-$ & $-$ &44992.77 & 797.00 & $-$ & $-$ &\\
MRCI~\cite{Langhoff_1988} & 1.650 & $-$ & $-$ & $-$ & 44106.00 & 813.00 & $-$ & $-$ & \\
MRCI\,+\,Q~\cite{Langhoff_1988} & 1.659& $-$ & $-$ & $-$ & 44224.00 & 797.00 & $-$ & $-$ &\\
Exp.NIST \cite{NIST, Herzberg_1979}& 1.6391 &$-$ & 0.56280 & 0.00651 & 44813.20 & 786.37 & 7.64 & 0.009 & \\
\hline
\multicolumn{10}{c}{$B^1\Sigma$}\\
\hline
\textbf{MRCI} & \textbf{1.667} & $-$ & \textbf{0.5441} & $-$ & \textbf{66891.64} & $-$& $-$& $-$\\
\textbf{MRCI\,+\,Q} & \textbf{1.613} & $-$ & \textbf{0.5812} & $-$ & \textbf{49640.25} & $-$ & $-$ & $-$ \\
MRCI~\cite{Well_2011} & 1.593 & $-$ & $-$ & $-$ & 56352.80 & $-$ & $-$ & $-$  \\
MRCI~\cite{Langhoff_1988} & 1.624 & $-$ &$-$ & $-$ & 54211.00 & $-$ & $-$ & $-$  \\
MRCI\,+\,Q~\cite{Langhoff_1988} & 1.632 & $-$ & $-$ & $-$ & 54258.00 & $-$ & $-$ & $-$  \\
Exp.~NIST \cite{NIST, Herzberg_1979} & 1.6151 &$-$ &0.57968 & $-$ & 54251.00 & $-$ & $-$ & $-$  \\
\end{tabular}
* Spectroscopic parameters deduced from a fit to all the measured frequencies.
\end{center}
\end{ruledtabular}
\end{table*}
\section{Theory and Methodology}
\label{section-2}
The electronic energy calculations for ground as well as for excited electronic states of the considered systems are performed using the MOLPRO~\cite{MOLPRO} quantum chemistry package. The spectroscopic constants and molecular properties have been calculated at MRCI level of theory in conjunction with the correlation-consistent polarized valence quadruple zeta (cc-PVQZ) basis sets. Further, in order to incorporate the unlinked quadruples contributions which actually improves the energies significantly and hence the potential energy surfaces, we have also considered the Davidson correction~\cite{MRCI+Q}.\\

To carry out MRCI calculations, a complete active space self consistent field (CASSCF) wave function is considered as the zeroth order wavefunction. The $C_{2v}$ subgroup of $C_{\infty v}$ point group has been used to perform all of the structure calculations. The $C_{2v}$ subgroup has four irreducible representations $A_1$, $B_1$, $B_2$, and $A_2$. The $A_1$ forms $\Sigma$, $B_1$ and $B_2$ together form $\Pi$, and $A_2$ forms $\Delta$ electronic states. Six orbitals for AlF and ten orbitals for AlCl form our frozen core. The complete active spaces (CAS), represented as ($A_1, B_1, B_2, A_2$), chosen for AlF and AlCl are (10, 4, 4, 1) and (12, 5, 5, 1), respectively. As a result, 13 orbitals form the CAS for electronic energy and property calculations for both the molecules. The potential energies, PDMs, and QMs are calculated with a step size of 1 \AA\,bond distance and in order to arrive at the equilibrium geometry, we have chosen a finer step size of $0.001$ \AA\,around the equilibrium point. The dissociation limits for AlF and AlCl molecules are $24$ \AA\,and $19.6$ \AA, respectively. The spectroscopic constants are computed using VIBROT program available in OpenMolcas~\cite{OPENMOLCAS} quantum chemistry software.\\

The traceless QM tensor, $\Theta_{\alpha\beta}$, can be defined as~\cite{Buckingham_1959},
\begin{eqnarray}
\Theta_{\alpha\beta}=\frac{1}{2}\,\sum_{i}e_i(3r_{i_\alpha}\,r_{i_\beta}\,-\,r_{i}^2\delta_{\alpha\beta}),
\end{eqnarray}
where $\alpha$ and $\beta$ represent the Cartesian components and summation index $i$ represents the number of charges.
By considering $z$ axis as the internuclear axis of the molecule, the $z$ component of QM, $\Theta_{zz}$, is related to two other diagonal components by the equation, 
\begin{eqnarray}
\Theta_{zz}\,=-(\,\Theta_{xx}\,+\,\Theta_{yy}).
\end{eqnarray}
Further, for linear molecules, $\Theta_{xx}\,=\,\Theta_{yy}$ and therefore,
\begin{eqnarray}
\label{QM}
\Theta_{zz}\,=\,-2\Theta_{xx}.
\end{eqnarray}

Using the PECs and PDM curves for ground and excited electronic states at MRCI level of theory, we have solved the vibrational Schr\"{o}dinger equation using the LEVEL~\cite{LEVEL} program to obtain the vibrational parameters: wavefunctions, energies, rotational constants, and TDMs between the vibrational levels and rotational parameters: energies, Einstein coefficients and Franck-Condon factors (FCFs). We have used cubic spline fitting for interpolation with suitable step size. Further, the relative energy separation and the TDMs between the vibrational levels are used to calculate the spontaneous and BBR induced transition rates at room temperature ($T = 300K$) using the following equations~\cite{Bala_2019, Bala_2018}:
\begin{eqnarray}{}  
\Gamma_{v, J}^{spon}\,=\,\sum\limits_{v^{''}, J^{''}}\Gamma^{emis}(v, J\,\rightarrow\,v^{''}, J^{''}) 
 \end{eqnarray}
and
 \begin{eqnarray}{} 
 \Gamma_{v, J}^{BBR}\,&=& \,\sum\limits_{v^{''}, J^{''}}\bar{n}(\omega)\,\Gamma^{emis}(v, J\,\rightarrow\,v^{''}, J^{''})\nonumber\\
 &+&\sum\limits_{v^{'}, J^{'}}\bar{n}(\omega)\,\Gamma^{abs}(v, J\,\rightarrow\,v^{'}, J^{'}),
  \end{eqnarray}
where the indices ($v^{''}, J^{''}$) and ($v^{'}, J^{'}$) denote the rovibrational levels with energy smaller and larger than that of ($v, J$) level, respectively.  
The average number of photons $\bar{n}(\omega)$ at frequency $\omega$ is given by the relation, \\
  \begin{eqnarray}{}
  \bar{n}(\omega)\,=\,\frac{1}{e^{(\hbar\omega/k_{B}T)}-1},
   \end{eqnarray}
where $\hbar\omega\,=\,\arrowvert E_{v, J}-E_{\tilde{v}, \tilde{J}}\arrowvert$ is the energy difference between the two rovibrational levels involved, with ($\tilde v, \tilde J$) being ($v^{''}, J^{''}$) for emission, and ($v^{'}, J^{'}$) for absorption. $k_{B}$ is the Boltzmann constant.
The emission and absorption rates are calculated using the equation,
 \begin{eqnarray}{}
 \Gamma^{emis\,or\,abs} (v, J\,\rightarrow\,v^{''}, J^{''}\,or\,v^{'}, J^{'})\,=\nonumber\\
 \,\frac{8\pi}{3\epsilon_0}\frac{1}{h c^3}\, \omega^3 (TDM_{v, J\rightarrow v^{''}, J^{''}\,or\,v^{'}, J^{'}})^2.
  \end{eqnarray}
  
  Finally, the total lifetimes ($\tau$) of the rovibrational states are obtained as,
\begin{eqnarray}{}
\tau\,=\,\frac{1}{\Gamma^{total}},
\end{eqnarray}
 where $\Gamma^{total}(\,=\,\Gamma^{spon}\,+\,\Gamma^{BBR}$) is the of sum of spontaneous and BBR 
 induced transition rates.\\
 
 The Franck-Condon factors, $q_{v',v"}$, are defined as the square of the matrix elements of the zeroth-power of the radial variable~\cite{LEVEL},
 \begin{eqnarray}
 q_{v',v"}\,=\,\mid\langle\Psi_{v',J'}|\Psi_{v",J"}\rangle|^2.
 \end{eqnarray}
\section{Results and Discussion}
Tables 1 through 6 and figure 1 to 4 present all of our calculated result, and their comparison with the existing data, wherever available. In all tables, the results computed in the current work are shown in bold fonts.
\label{section-3}
\subsection{AlF}
\subsubsection{Electronic properties: ground and low-lying excited states} 
The PECs for the ground and low\,-\,lying excited states with $\Sigma$ and $\Pi$ symmetries at MRCI+Q  level of theory are shown in Figure~\ref{fig:FIG1}(a). The spectroscopic constants for the ground electronic state calculated in the current work along with the results available in the literature are given in Table~\ref{tab:table1}. The results for spectroscopic parameters obtained in this work compares well with the existing values. Except for $\omega_e x_e$, the difference between our results for ground spectroscopic constants from that reported in the recent work by Zhang {\it et al.}~\cite{Zhang_2021} is less than 1\% at the same level of theory. Although, $\omega_e x_e$ is extracted from the PECs along with the other spectroscopic parameters via fitting, it is quite sensitive to the number of data points used for fitting. The relative difference between our results obtained using MRCI+Q theory from those given in  Ref.~\cite{Gutsev_1999} at CCSD(T) level is: $\Delta (R_e)\,=\,0.44\%$, $\Delta (D_e)\,=\,0.17\%$, $\Delta(B_e)\,=\,1\%$ and $\Delta (\omega_e)\,=\,0.6\%$ for the ground electronic state. Comparing our values of $R_e$, $D_e$, and $B_e$  to the experimental values~\cite{Hedderich_1993, Lide_1965}, we observe that the  errors are 0.7\%, 1.1\% and 1.7\%, respectively. 
\begin{table}[htbp]
\caption{\label{tab:table3}
Molecular properties (electric dipole and electric quadrupole moments denoted as $\mu_e$ and $\Theta_{zz}$, respectively) for the ground and excited states of AlF molecule.}
\begin{ruledtabular}
\begin{tabular}{cccccccc}
State & Method &$\mu_e$ (Debye)  & $\Theta_{zz}$ (au)\\
\hline
$X^1\Sigma$ &\textbf{SCF} & \textbf{-1.285} & \textbf{-5.929}\\
         &\textbf{CASSCF} & \textbf{-1.434} & \textbf{-5.379}\\
        &\textbf{MRCI} & \textbf{-1.425} &\textbf{-5.277}\\
        & MRCI \cite{Zhang_2021}&-1.079&$-$\\
        & CCSD(T) \cite{Woon_2009} & 1.540 & $-$\\
        & SCF \cite{Gutsev_1999}& 1.357 & -4.674\\
        & CCSD(T) \cite{Gutsev_1999}& 1.486 & -3.766\\
        & ACPF\cite{Yousefi_2018}& 1.407 & $-$\\
        &Exp.\cite{Lide_1965}& 1.53$\pm$ 0.10& $-$\\
\hline
$a^3\Pi$ & \textbf{MRCI} & \textbf{-1.770} & \textbf{-2.794}\\
$A^1\Pi$ & \textbf{MRCI} & \textbf{-1.546} & \textbf{-1.287}\\
         & MRCI\cite{Zhang_2021}& 1.497& $-$\\
$b^3\Sigma$ & \textbf{MRCI} & -\textbf{1.699} & \textbf{-9.397}\\
$B^1\Sigma$& \textbf{MRCI} & \textbf{-1.811} & \textbf{-5.295}\\
\end{tabular}
\end{ruledtabular}
\end{table}
\begin{figure*}[t]
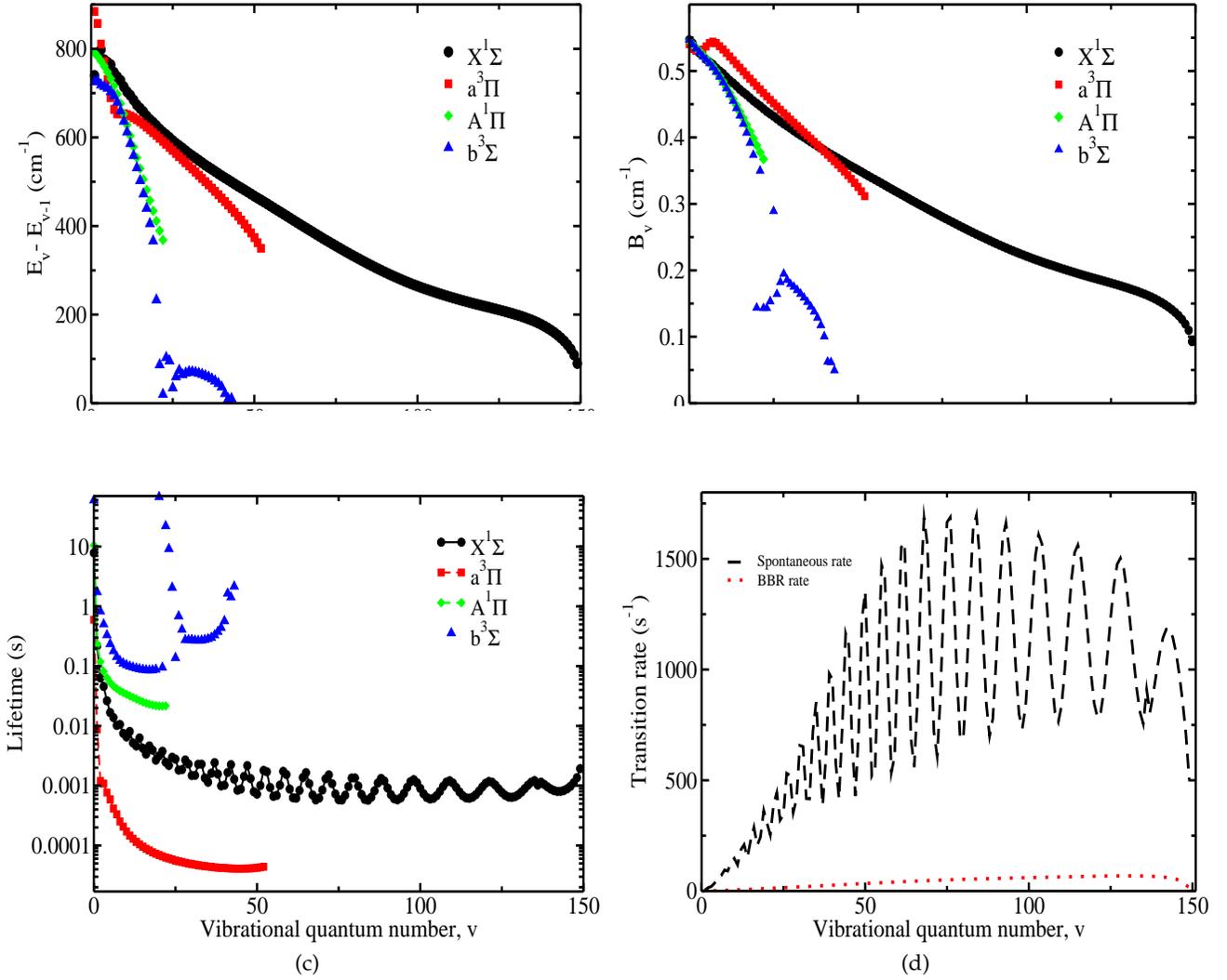

\begin{tabular}{cc}
\includegraphics[width=8.3cm, height=6.5cm]{AlF_vib_energy.eps}&\includegraphics[width=8.5cm, height=6.5cm]{AlF_rot_const.eps}\\ 
(a)&(b)\\
\includegraphics[width=8.5cm, height=6.5cm]{AlF_vib_lifetime.eps}\,\,&\includegraphics[width=8.4cm, height=6.5cm]{AlF_GS_transition_rate.eps}\\ 
(c)&(d)\\
\end{tabular}
\caption{\label{fig:FIG2} (a) Relative energy spacings between the vibrational levels (b) rotational constants (c) lifetimes and (d) transition rates for the vibrational levels of ground and excited states of AlF molecule at MRCI/QZ level of theory.}
\end{figure*}
Our value of $D_e = 7.022$\,eV at MRCI+Q level matches well within the error bars with the measured value of $6.894\,\pm\,0.130$ reported in Ref.~\cite{Murad_1966}. We could not find any value for $\omega_ey_e$ in the literature for comparison. The value of $\alpha_e$ reported in this work at MRCI level is very close to the existing theoretical and experimental results.\\

Table~\ref{tab:table2} gives the comparison of results for spectroscopic parameters of low-lying excited states obtained in this work with the literature. Except for $A^1\Pi$ electronic state, dissociation energies are not available for other excited states studied in this work. Our value of $D_e$ for $A^1\Pi$ state is in excellent agreement with the only result available in Ref.~\cite{Zhang_2021}. The MRCI\,+\,Q value of $\omega_ex_e\,=\,5.70$\,cm$^{-1}$ for $A^1\Pi$ state is very close to the experimental value of $5.99$\,cm$^{-1}$~\cite{Herzberg_1979}, while for $a^3\Pi$ ($B^1\Sigma$) state, our value is larger by 1.53\,cm$^{-1}$ (1.17\,cm$^{-1}$) from the experimental data. The $\omega_ey_e$ value for $a^3\Pi$ state is of the same order as reported in Ref.~\cite{Walter_2022}, whereas the $\omega_ey_e$ values for $A^1\Pi$ and $b^3\Sigma$ states are one order of magnitude higher than those given in NIST database~\cite{NIST, Herzberg_1979}. The computed results for $\alpha_e$ in the present work are of the same order as the data given in Refs.~\cite{Walter_2022, NIST, Herzberg_1979}. For the other excited state parameters, the relative difference between the values reported in our work and the experimental data available in literature lies in the range 0.9\,-\,1.3\% for $R_e$, 1.6\,-\,2.2\% for $B_e$, 0.2\,-\,2.7\% for $T_e$, and 0.05\,-\,0.6\% for $\omega_e$.\\
\begin{table*}[htbp]
\caption{\label{tab:table4}
Spectroscopic constants for the ground electronic state of AlCl molecule.}
\begin{ruledtabular}
\begin{tabular}{ccccccccc}
Method & $R_e$ (\AA) & $D_e (eV)$ & $B_e$ (cm$^{-1}$) & $\alpha_e$ (cm$^{-1}$) & $\omega_e$ (cm$^{-1}$) & $\omega_ex_e$ (cm$^{-1}$) & $\omega_e y_e$ (cm$^{-1}$) \\
\hline
\textbf{SCF} & \textbf{2.149} & \textbf{6.590} & \textbf{0.2393} & \textbf{0.0016} & \textbf{476.57} & \textbf{2.19} & \textbf{0.0051}\\
\textbf{CASSCF} & \textbf{2.141} & \textbf{7.356} & \textbf{0.2314} & \textbf{0.0007} & \textbf{400.55} & \textbf{7.46} & \textbf{0.1387}\\
\textbf{MRCI} & \textbf{2.143} & \textbf{5.539} & \textbf{0.2400} & \textbf{0.0013} & \textbf{478.61} & \textbf{1.10} & \textbf{0.0115}\\
\textbf{MRCI\,+\,Q} & \textbf{2.145} & \textbf{5.274} & \textbf{0.2403} & \textbf{0.0015} & \textbf{483.47} & \textbf{2.20} & \textbf{0.0066}\\
\hline
\textbf{Published works}\\
CASSCF \cite{Langhoff_1988} & 2.157 & 4.360 & $-$ & $-$ & 470.00 & $-$ & $-$\\
MRCI\cite{Langhoff_1988} & 2.142 & 5.120 & $-$ & $-$ & 496.00 & $-$ & $-$\\
MRCI\,+\,Q \cite{Langhoff_1988}&2.140 & 5.270 & $-$ & $-$ & 500.00 & $-$ & $-$ \\
ACPF \cite{Yousefi_2018}& 2.128 & $-$ & 0.2441 & 0.0016 & 484.81 & 2.07 & $-$\\
MRCI \cite{Yang_2016} & 2.145 & 5.220 & 0.2406 & $-$ & 478.36 & 1.95 & $-$\\
MRCI\,+\,Q \cite{Xu_2020} & 2.137 & 5.262 & 0.2425 & $-$ & 481.83 & 2.01 & $-$\\
MRCI \cite{Wan_2016} & 2.137 & 5.214 & 0.2408 & $-$ & 478.13 & $-$ & $-$\\
 MRCI\,+\,Q \cite{Daniel_2021} & 2.137 & $-$ & $-$ & $-$ & $-$ & $-$ & $-$\\
 CCSD(T) \cite{Woon_2009} & 2.159 & $-$ & 0.2360 & $-$& $-$ & $-$ & $-$\\
 MRCI\,+\,Q \cite{Brites_2008} & 2.140 & 5.250 & 0.2418 & 0.0017 & 484.50 & 6.47 & 0.9800\\ 
Exp.\cite{Lide_1965}& 2.12983 & $-$ & 0.24393 & 0.00160 & $-$ & $-$ &$-$\\
Exp. \cite{Hedderich_1993} & 2.130 & 5.120 & $-$ & $-$ & $-$ & $-$ & $-$ \\
Exp. \cite{Ram_1982}&$-$ & 5.25 $\pm$ 0.01 & 0.2439 & 0.00161 & $-$ & $-$ & $-$\\
NIST \cite{NIST}& 2.130113 & $-$ &  0.24393102 & 0.00161113 & 481.3 &  1.95 &$-$ \\
\end{tabular}
\end{ruledtabular}
\end{table*}

The $B^1\Sigma$ state has a minimum energy at $R_e\,=\,1.613$\,\AA\ and its energy attains its maximum value at about $1.8$\,\AA\,, after which it starts decreasing. This trend in the PEC is same as that described in Ref.~\cite{Well_2011}. It can be seen from Figure~\ref{fig:FIG1}(a) that the $B^1\Sigma$ state has a shallow minimum of $1098.8$\,cm$^{-1}$, which renders the computation of other constants via fitting of PEC meaningless. The values of equilibrium bond length and rotational constant of this state at MRCI\,+\,Q level of theory agrees with the values in  NIST database. However there is a noticeable difference in the values of transition energies.\\

The computed first\,-\,order molecular properties at the equilibrium bond lengths are given in Table~\ref{tab:table3}. The variation of PDM as a function of internuclear distance, for all the electronic states, is shown in Figure~\ref{fig:FIG1}(b). The $R$ variation of PDM for the ground state is linear almost upto $9$\,\AA\,and then it drops to zero abruptly. The linear behaviour of the PDM curve signifies the ionic character of the molecule and the abrupt change indicates the change from ionic to covalent character of the dipole~\cite{Well_2011}. Further, it has been found that using state\,-\,averaging over states of symmetry same as that of the ground state, the PDM for the ground state goes to zero after $4$\,\AA. A very similar behaviour of PDM curve of this system has been reported in Ref.~\cite{Zhang_2021}, though they have not mentioned about the usage of state\,-\,averaged procedure  in their work. It has been observed that the values of PDMs are almost zero at the dissociation limit for all electronic states, which means that the molecule dissociates into its constituent neutral fragments. \\

The variation of the $z$\,-\,component of QM with internuclear distance is given in Figure~\ref{fig:FIG1}(c). Similar to PDM, the QM value also drops and attains a small constant value of about 2.16\,au (-3.15\,au) after $4$\,\AA\,($9$\,\AA) with (without) state\,-\,average procedure. The QM values for the excited states decrease quadratically upto 2\AA\,bond distance for $a^3\Pi$, $A^1\Pi$, and $b^3\Sigma$ states, while for $B^1\Sigma$ state, this behaviour is noticed upto the bond length of 1.7\,\AA\,as shown in the subplot of Figure~\ref{fig:FIG1}(c). The quadrupole moment relation for linear diatomic molecules given by eqn.~(\ref{QM}) is satisfied for ground $^1\Sigma$ states upto $9.5$\,\AA\, (4\,\AA) bond distance with (without) state average procedure. The TDM curves from ground to singlet excited states are plotted in Figure~\ref{fig:FIG1}(d). The TDM goes to zero after $5$\,\AA\,for $X^1\Sigma \rightarrow A^1\Pi$ transition and for $X^1\Sigma \rightarrow B^1\Sigma$, its value is zero $2$\,\AA\,onwards due to the fact that these transitions are forbidden in atomic limits. 
\subsubsection{Vibrational and rotational parameters} 
We have found $150$ bound vibrational states for $X^1\Sigma$, $53$ for $a^3\Pi$, $23$ for $A^1\Pi$, and $44$ for $b^3\Sigma$ electronic states.  Figure~\ref{fig:FIG2}(a) and Figure~\ref{fig:FIG2}(b) presents the behaviour of relative energy spacings and vibrational\,-\,rotational constants against the vibrational quantum number, respectively. The relative energy separation between the vibrational levels and the vibrational\,-\,rotational constants decreases with the increase in vibrational quantum number due to anharmonic effects. We have also calculated the first six centrifugal distortion constants  {\it viz.} $D_v, H_v, L_v, M_v, N_v$ and $O_v$ for the bound states of AlF. These results are presented in Tables S1 to S4 of supplementary material. The PDMs for the vibrational levels (with $J\,=\,0$) of ground and excited electronic states are tabulated in Supplementary Tables S5 and S6, respectively. We have obtained the following values of the electric dipole moments $1.421$ D for $X^1\Sigma$, $1.567$ D for $A^1\Pi$ and $1.241$ D for $a^3\Pi$ in their lowest ro\,-\,vibronic levels. 
The only measured values of dipole moments for $X^1\Sigma$, $A^1\Pi$ and $a^3\Pi$ corresponding to $v\,=\,0$ vibrational levels are $1.515\,\pm\,0.004$ D, $1.45\,\pm\,0.02$ D, and $1.780\,\pm\,0.003$ D, respectively~\cite{Truppe_2019}. Our results differ from those reported in Ref.~\cite{Truppe_2019} by $0.094$ D for $X^1\Sigma$, $-0.117$ D for $A^1\Pi$ and $0.539$ D for $a^3\Pi$. 
\begin{figure*}[ht]
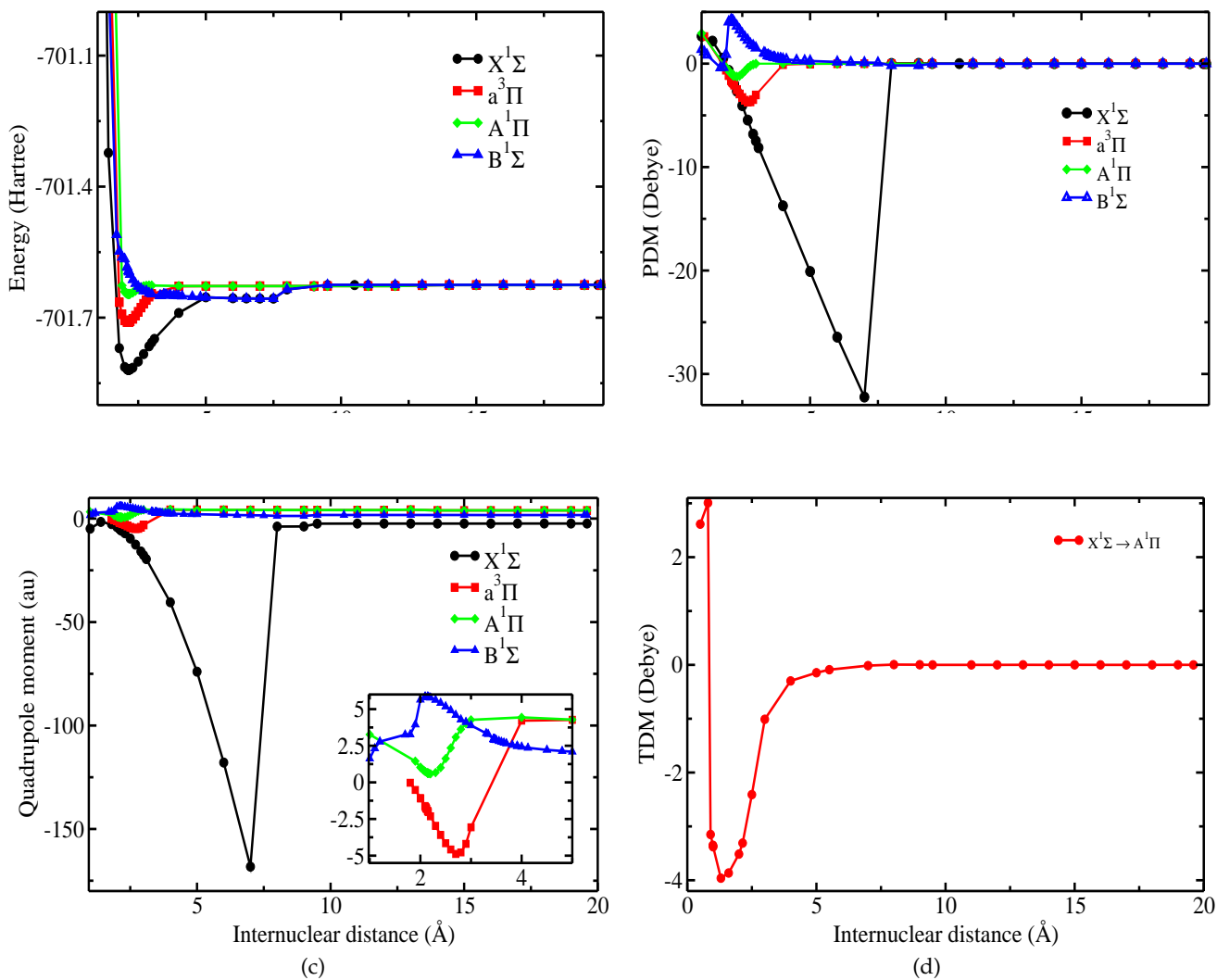

\begin{tabular}{cc}
\includegraphics[width=8.6cm, height=6.5cm]{AlCl_MRCIQ_PEC_corrected.eps}\,\,\,&\includegraphics[width=8.2cm, height=6.5cm]{AlCl_GS_ES_PDM_MRCI.eps}\\
(a)&(b)\\
\hspace{0.2cm}
\includegraphics[width=8.5cm, height=6.5cm]{AlCl_corrected_QM_GS_ES.eps}\,\,\,\,\,&\includegraphics[width=8.3cm, height=6.5cm]{AlCl_TDM_GS_sing_pi.eps}\\
(c)&(d)\\
\end{tabular}
\caption{\label{fig:FIG3} (a) Potential energy curves (b) permanent dipole moment curves (c) quadrupole moment curves for the ground and low-lying excited electronic states and (d) transition dipole moment curves from ground to singlet excited electronic states of the AlCl molecule.}
\end{figure*}
Yousefi {\it et al.}~\cite{Yousefi_2018} have obtained the PECs for the ground states of AlF and AlCl using direct fitting of rotational and ro\,-\,vibrational frequencies together with the dissociation energies taken from the literature~\cite{Hedderich_1993}. Further, they have studied the vibration\,-\,rotation line lists upto $v\,=\,11$ and maximum number of rotational levels, $J_{max}\,=\,200$ for electronic ground states of these molecules. They have reported $\mu_v\,=\,1.439$ D for the lowest ro\,-\,vibrational level of ground electronic state. The difference between our value and that reported in their work is less than 1.3\%. \\

The variation of total lifetimes against the vibrational quantum number, $v$ is shown in Figure~\ref{fig:FIG2}(c). The lifetime curve of ground electronic state of AlF molecule shows several minor oscillations similar to CaLi$^+$ system in Ref.~\cite{Bala_2019}. This is due to the oscillations in the spontaneous transition rate as shown in Figure~\ref{fig:FIG2} (d) and also observed for CaLi$^+$ ionic system in Ref.~\cite{Bala_2019}. The calculated lifetimes for the lowest vibrational\,-\,rotational states ($v\,=\,0$ and $J\,=\,0$) of $X^1\Sigma$, $a^3\Pi$, $A^1\Pi$, and $b^3\Sigma$ are $7.8$\,s, $0.59$\,s, $10$\,s, and $59$\,s, respectively.\\

Further, we have computed the rotational parameters: relative energies, Einstein coefficients, FCFs, and TDMs. The rotational transitions corresponding to $\Delta J\,=\,0$, $\Delta J\,=\,+1$ and $\Delta J\,=\,-1$ give the set of lines called $Q$, $R$, and $P$\,-\,branches, respectively. We have performed the calculations of rotational parameters for all $P$, $Q$ and $R$ branches considering $J_{max}$ is equal to 20 in each vibrational level $v\,=\,0$ to $5$ in ground as well as excited electronic states. All parameters for the rotational transitions between the vibrational levels of individual electronic states are given in supplementary Tables S7 to S10. The Einstein coefficients between the rotational levels of different vibrational levels in each electronic state are plotted in Figures 1 to 4 of supplementary file PS1. 
\begin{table*}[htbp]
\begin{ruledtabular}
\begin{center}
\caption{\label{tab:table5}
Spectroscopic constants for the excited electronic states of the AlCl molecule. }
\begin{tabular}{cccccccccc}
 Method & $R_e$ (\AA)&$D_e$ (eV) & $B_e$ (cm$^{-1}$)& $\alpha_e$ (cm$^{-1}$) & $T_e$ (cm$^{-1}$)&$\omega_e$ (cm$^{-1}$)&$\omega_e x_e$ (cm$^{-1}$) & $\omega_e y_e$ (cm$^{-1}$) & Ref.\\
\hline
\multicolumn{10}{c}{$a^3\Pi$} \\
\hline
\textbf{MRCI} & \textbf{2.112} & \textbf{2.591} & \textbf{0.2449} & \textbf{0.0006} & \textbf{24014.24}  & \textbf{500.68} & \textbf{0.25} & \textbf{0.0388} \\
\textbf{MRCI\,+\,Q} & \textbf{2.114} & \textbf{2.320} & \textbf{0.2453} & \textbf{0.0008} & \textbf{23970.88} & \textbf{509.07} & \textbf{1.40} & \textbf{0.0263} \\
NIST \cite{NIST} & 2.100 & $-$ & 0.2500 & 0.0020 & 24658.00, & 524.35 & 2.18 & $-$  \\
&&&&&24593.84, 24528\\
 MRCI\,+\,Q~\cite{Brites_2008} & 2.107 & $-$ & 0.2494 &  0.0015& 24519.25& 519.10 & 0.52& 0.2300  \\
MRCI~\cite{Langhoff_1988} &2.110 & $-$ & $-$ & $-$& 23819.00 & 530.00 & $-$ & $-$ \\ 
 MRCI\,+\,Q~\cite{Langhoff_1988} & 2.114 &$-$ & $-$& $-$&24194.00& 519.00&$-$ & $-$ \\
MRCI~\cite{Yang_2016} & 2.112 & 2.240 & 0.2481 & $-$ & 24057.00 & 523.36 & 2.61 & $-$ \\
 MRCI\,+\,Q~\cite{Xu_2020} & 2.104 & 2.261 & 0.2501 & $-$ & 24223.52 & 527.37 & 2.65 & $-$  \\
 MRCI~\cite{Wan_2016} & 2.105 & 2.246 & 0.2483 &$-$ & 23959.87 & 525.68 & $-$ & $-$  \\
 Exp.~\cite{Ram_1982} & $-$ & $-$ & 0.2524 & 0.0011 & 24793.10 &$-$ & $-$ & $-$  \\
\hline
\multicolumn{10}{c}{$A^1\Pi$}\\
\hline
\textbf{MRCI} & \textbf{2.146} & \textbf{0.730} & \textbf{0.2409} & \textbf{0.0029} & \textbf{38955.52} & \textbf{445.76} & \textbf{9.15} & \textbf{0.0532}\\
\textbf{MRCI\,+\,Q} & \textbf{2.146} & \textbf{0.568} & \textbf{0.2394} & \textbf{0.0016} & \textbf{38015.98} & \textbf{445.62} & \textbf{7.15} & \textbf{0.0361} \\
 MRCI\,+\,Q~\cite{Brites_2008}& 2.132 & $-$ & 0.2435 & 0.0027 & 38795.26 & 453.00 & 8.03   & 0.4400\\ 
NIST \cite{NIST} & 2.060 & $-$ & 0.2590 & 0.0060 & 38254.00 & 449.96 & 4.37 & 0.2160 \\
MRCI~\cite{Langhoff_1988} & 2.138 & $-$ & $-$ &$-$ & 39630.00 & 464.00 &$-$& $-$\\
 MRCI\,+\,Q~\cite{Langhoff_1988} & 2.152 & $-$ & $-$ &$-$ & 38453.00 & 449.00 &$-$& $-$&\\
 MRCI~\cite{Yang_2016} & 2.142 & 0.530 & 0.2412 & $-$ & 38303.00  & 471.83 & 9.61 & $-$ \\
 MRCI\,+\,Q~\cite{Xu_2020}& 2.132 & 0.496 & 0.2435 & $-$ & 38436.37 & 453.43 & 8.48 & $-$ \\
 MRCI~\cite{Wan_2016} & 2.133 & 0.544 & 0.2397 & $-$& 38223.98 & 454.24&$-$& \\
 MRCI\,+\,Q~\cite{Daniel_2021}& 2.133 & $-$ & 0.24313 & 0.0030 & 38253.72 & 446.26 & 5.04 & $-$ &\\
 Exp.~\cite{Daniel_2021} & 2.1220 & $-$ & 0.24535 (2) & 0.002652(7) &  38253.22 (2)& 452.54 (5) & 5.61 (3) & $-$  \\
 Exp.~\cite{Ram_1982} & $-$ & $-$ & 0.24537 & 0.0025 & 38237.00 & $-$ & $-$ & $-$  \\ 
\hline
\multicolumn{10}{c}{$B^1\Sigma$}\\
\hline
\textbf{MRCI} & $-$ & $-$ & $-$ &$-$ & \textbf{56763.11} & $-$ & $-$  & $-$ \\
 \textbf{MRCI\,+\,Q} & $-$ & $-$ & $-$ &$-$ & \textbf{49649.42} & $-$ & $-$  & $-$\\
 MRCI\,+Q~\cite{Brites_2008} & $-$ & $-$ & $-$ & & 54119.79 & $-$ & $-$  & $-$ \\
 MRCI~\cite{Langhoff_1988} & 2.053 & $-$ & $-$ & $-$ & 53003.00 &  496.00 & $-$  & $-$ \\
MRCI\,+\,Q~\cite{Langhoff_1988} & 2.069 & $-$ & $-$ & $-$ & 52800.00 &  468.00 & $-$  & $-$  \\
\end{tabular}
\end{center}
\end{ruledtabular}
\end{table*}
These results for rotational levels, particularly for the excited states, may useful for the future line list observations in astrophysical environments. 
\subsection{AlCl}
\subsubsection{Electronic properties: ground and low-lying excited states} 
The potential energies for the electronic states of AlCl molecular system studied in this work at MRCI\,+\,Q level of theory, are plotted against the internuclear distances in Figure~\ref{fig:FIG3}(a). The comparison of spectroscopic constants computed in the current work for ground and excited electronic states together with the available results in the literature are collected in Table~\ref{tab:table4} and Table~\ref{tab:table5}, respectively.\\

\begin{table}[htbp]
\caption{\label{tab:table6}
Molecular properties (electric dipole and electric quadrupole moments denoted as $\mu_e$ and $\Theta_{zz}$, respectively) for the electronic ground and excited states of AlCl molecule.}
\begin{ruledtabular}
\begin{tabular}{cccccccc}
State & Method &$\mu_e (Debye)$  & $\Theta_{zz}$ (au)\\
\hline
$X^1\Sigma$ & \textbf{SCF} & \textbf{-1.604} & \textbf{-6.387}\\
         &\textbf{CASSCF} & \textbf{-1.549} & \textbf{-5.379}\\
        & \textbf{MRCI} & \textbf{-1.589} & \textbf{-5.377}\\
        & ACPF \cite{Yousefi_2018}& 1.594&$-$\\
        & MRCI \cite{Xu_2020}& 0.900& $-$\\
        & MRCI \cite{Wan_2016} & -1.308 & $-$\\
\hline
$a^3\Pi$ & \textbf{MRCI} & \textbf{-1.669} & \textbf{-1.749}\\
       &MRCI\cite{Xu_2020}&1.770& $-$\\
       & MRCI \cite{Wan_2016} & -1.760 & $-$\\
$A^1\Pi$ & \textbf{MRCI} & \textbf{-1.084} & \textbf{0.640}\\
         & MRCI\cite{Xu_2020} &1.337& $-$\\
         & MRCI \cite{Wan_2016}& -0.918 & $-$\\
\end{tabular}
\end{ruledtabular}
\end{table}
The computed results at MRCI\,+\,Q level of theory for $R_e$, $D_e$, $B_e$, and $\alpha_e$ of the ground electronic state show very good agreement with the existing theoretical and experimental data~\cite{Lide_1965, Hedderich_1993, Ram_1982}. 
The values of diatomic constants ($R_e$, $D_e$, $B_e$, and $\omega_e$) reported in this work do not differ by more than 0.92\% from those reported recently in Ref.~\cite{Xu_2020} at the same level of correlation. Further, the value of $\omega_ex_e$ which depends on the other diatomic constants shows an appreciable difference of about 9.4\% between the two works. In Ref.~\cite{Brites_2008}, the authors have reported all the spectroscopic parameters and the results for $R_e$, $D_e$, $B_e$, $\alpha_e$ and $\omega_e$ show excellent agreement with those reported in the current work. However, the value of anharmonic constant, $\omega_ex_e\,=\,6.47$\,cm$^{-1}$ reported in their work is large as compared to any other value reported in the literature including ours and this difference lies in the range 4.27 to 4.52 cm$^{-1}$. Thus, we refrain from comparing our value for higher anharmonic constant, $\omega_ey_e$, with that reported in Ref.~\cite{Brites_2008}.\\

The computed spectroscopic constants: $R_e$, $D_e$, $B_e$, $T_e$ and $\omega_e$ in the present work for excited states compares well with the published results. The vertical transition energy for $a^3\Pi$ ($A^1\Pi$) excited state at MRCI\,+\,Q level matches with the experimental result to within 3.3\% (0.6\%)~\cite{Ram_1982, Daniel_2021}. The existing values of $\omega_ex_e$ for $a^3\Pi$ and $A^1\Pi$ states lie in the range 0.52 to 2.6540 cm$^{-1}$ and 4.37 to 9.61 cm$^{-1}$, respectively and our calculated values lies in between them. The computed $\omega_ey_e$ values for the excited state are one order of magnitude smaller than the available results~\cite{NIST, Brites_2008}.\\

There is a disagreement between the results of Langhoff {\it{et al.}}~\cite{Langhoff_1988} and Brites~{\it et al.} for the first excited $^1\Sigma$ state. In the former work, they have reported equilibrium bond length, harmonic frequency and transition energy, which 
clearly gives the idea of attractive the nature of the state while, the latter work shows the repulsive nature of the PEC. In the present work, we have found that the $B^1\Sigma$ state to be repulsive. Further, there is a mismatch between the existing results of transition energies for this state~\cite{Brites_2008,Langhoff_1988}. The value of $T_e$ reported in Ref.~\cite{Langhoff_1988} at MRCI\,+\,Q level of theory is smaller by 1319.8 cm$^{-1}$ from that reported in \cite{Brites_2008}.  Our value of $T_e$ using MRCI (MRCI\,+\,Q) method is larger (smaller) by 3760.11 cm$^{-1}$ (3150.58 cm$^{-1}$ ) from that reported in Ref.~\cite{Langhoff_1988}  at a similar level of theory.\\

The behaviour of PDMs and $z$\,-\,component of QMs with the bond distances are shown in Figure~\ref{fig:FIG3}(b) and Figure~\ref{fig:FIG3}(c), respectively.
The results for the molecular properties of AlCl are presented in Table~\ref{tab:table6}. The PDM varies linearly upto approximately $7$\,\AA\,for $X^1\Sigma$, $3$\,\AA\,for $a^3\Pi$ and $2.4$\,\AA\,for $A^1\Pi$ state. The PDM curve for excited singlet sigma state shows the maximum value of $4.236$\, D at  $2.09$\AA\, bond distance. The absolute values of PDMs for  $X^1\Sigma$, $a^3\Pi$, $A^1\Pi$, and $B^1\Sigma$  states at the equilibrium bond lengths are  $1.589$\,D, $1.669$\,D, $1.084$\,D, and $10.314$\,D, respectively. Our result for the PDM of electronic ground state is very close to that reported in Ref.~\cite{Yousefi_2018} and the difference between the two works is only $0.32$\%. However, our value of PDM is larger by 0.689\,D and 0.281\,D from that reported by Xu {\it et al.}~\cite{Xu_2020} and Wan {\it et al.} ~\cite{Wan_2016}, respectively at the same level of correlation.  The maximum difference between the current value of PDM from those reported in literature for $a^3\Pi$ ($A^1\Pi$) state is $0.1$\,D (0.25\,D). 
\begin{figure}[H]
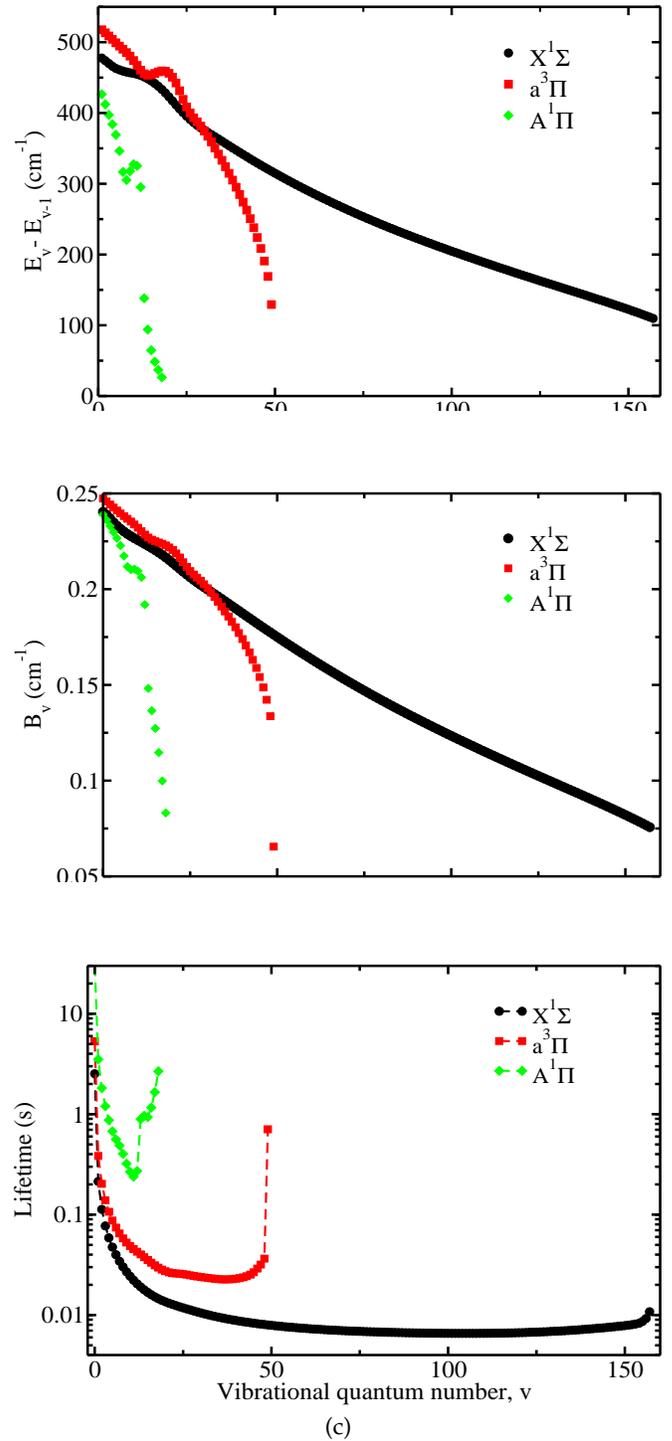

\addtolength{\tabcolsep}{0.1cm}
\begin{tabular}{c}
\includegraphics[width=\columnwidth,height=5.9cm]{AlCl_vib_energy.eps}\\
(a)\\
\includegraphics[width=\columnwidth,height=5.9cm]{AlCl_rot_const.eps}\\
(b)\\
\includegraphics[width=\columnwidth,height=5.9cm]{AlCl_lifetime.eps}\\
(c)\\
\end{tabular}
\caption{\label{fig:FIG4} (a) Relative energy spacing (b) rotational constants (c) lifetimes of the vibrational levels of ground and excited states of AlCl molecule at MRCI/QZ level of theory.}
\end{figure}

The results of the QMs for this molecular system is reported for the first time in this work to the best of our knowledge. The knowledge of multipole moments is useful for better understanding the internuclear forces/electrostatic bonding~\cite{Buckingham_1959, Glaser_2000}. Figure~\ref{fig:FIG3}(d) shows the TDM curve from $X^1\Sigma$ to $A^1\Pi$ state. The value of TDM reaches to a minimum of $-3.96$\,D at $1.3$\,\AA\,and then drops to almost zero at $8$\,\AA. The TDM value for $X^1\Sigma \rightarrow A^1\Pi$ transition at $R_e\, (2.143)$\AA\,is found to be 3.310\,D,  which is small than $3.392$\,D at $R_e\,=\,2.137$\AA\,reported in Ref.~\cite{Wan_2016} by 2.4\%. 
\subsubsection{Vibrational and rotational parameters} 
We have found $158$,  $50$ and $19$ bound vibrational states for $X^1\Sigma$, $a^3\Pi$ and $A^1\Pi$ electronic states, respectively while Xu {\it et al.} have reported 159 vibrational levels for $X^1\Sigma$, $53$ for $a^3\Pi$, and 10 for $A^1\Pi$  state. The difference in number of vibrational states in both the works is mainly for the  $A^1\Pi$ state and arises due to the difference in the depths of potential energy surfaces.  The value of potential depth for $A^1\Pi$ state reported in Ref.~\cite{Xu_2020} is smaller than any other calculated value in the literature including ours. The relatively small value of dissociation energy accounts for lesser number of vibrational levels in Ref.~\cite{Xu_2020}.  Similar to AlF, the vibrational energy spacings and vibrational\,-\,rotational constants decrease with increase in vibrational quantum number as shown in Figure~\ref{fig:FIG4}(a) and Figure~\ref{fig:FIG4}(b). The centrifugal distortion constants for the vibrational levels of electronic states are given in supplementary tables S11 to S13. The difference in zero point energies ($E_{v, J\,=\,0}$) calculated in our work and that reported in Ref.~\cite{Xu_2020} is 0.06\% for $X^1\Sigma$, 0.99\% for $a^3\Pi$, and $3.25\%$ for $A^1\Pi$ state. Supplementary Tables S14 and S15 present the PDMs of ro\,-\,vibrational levels of ground and excited electronic states, respectively. The PDM value of $-1.621$\,D for the lowest ro\,-\,vibrational level of $X^1\Sigma$ state obtained in the current work differs by 0.5\% from that reported in Ref.~\cite{Yousefi_2018}. The total lifetimes of vibrational levels (with rotational quantum number $J\,=\,0$)  of the studied electronic states decrease initially and then begin to increase as shown in Figure~\ref{fig:FIG4}(c). 
We have obtained $2.5$\,s, $5.3$\,s, and $30$\,s lifetimes for $v\,=\,0, J\,=\,0$ level of $X^1\Sigma$, $a^3\Pi$, and $A^1\Pi$, respectively. The longer lifetimes for the higher vibrational states have also been reported in Ref.~\cite{Zemke_2004}. The results for $P$, $Q$ and $R$ branches due to the rotational transitions between the vibrational levels are presented in Figures 5 to 7 of supplementary file PS1.
\section{Conclusion}
\label{section-4}
In summary, we have carried out extensive theoretical electronic, vibrational and rotational structure calculations that include a wide range of properties, on the AlF and AlCl molecular systems. The ground and low\,-\,lying excited states of both the systems have been studied at MRCI+Q level of theory using QZ basis sets. Our findings for the diatomic constants (spectroscopic parameters) agree well with existing results. The calculations performed in this work report some results for the first time, and also resolve discrepancies in some existing calculations. We have also studied the vibrational levels and evaluated their wavefunctions, energies, dipole moments, spontaneous as well as BBR-induced transition rates, and lifetimes. Further, we extended this work to the rotational levels corresponding to vibrational levels of all electronic states. We have reported transitions between the rotational levels for P, Q, and R branches in the vibrational levels of all the electronic states. We expect that the $ab$ $initio$ results that are reported in this work for diatomic constants, molecular properties as well as vibrational and rotational parameters of ground and excited electronic states, would be useful for future spectroscopic studies including potential applications ranging from ultracold molecules to astrophysics.
\begin{acknowledgments}
All the structure calculations using MOLPRO program have been performed at the computation facility available at Indian Association for the Cultivation of Science (IACS), Kolkata. 
\end{acknowledgments}

\bibliography{AlF_AlCl}

\end{document}